\documentclass{article}
\usepackage{epsfig}
\def\ha{H_1}
\def\hb{H_2}

\begin{document}

\title{Four-dimensional lattice results on the MSSM electroweak phase 
transition$^{\dagger}$       
}

\author{F.~Csikor$^{\rm a}$, Z.~Fodor$^{\rm a}$, P.~Heged\"us$^{\rm a}$,
A.~Jakov\'ac$^{\rm b}$,\\
S.~D.~Katz$^{\rm a}$, A.~Pir\'oth$^{\rm a}$\\
$^{\rm a}$ Institute for Theoretical Physics, E\"otv\"os
University, \\ H-1117, P\'azm\'any P. 1A, Budapest, Hungary\\
$^{\rm b}$ Department of Theoretical Physics, Technical
University, Budapest, \\ H-1521, Budapest, Hungary
}

\maketitle
\abstract{
We present the results of our large scale 4-dimensional (4d)
lattice simulations for the  MSSM electroweak phase transition (EWPT). 
We carried out infinite volume
and continuum limit extrapolations and found a transition whose 
strength agrees well with perturbation theory. We determined the 
properties of the bubble wall that are important for a successful
baryogenesis.
}

%---------------------------------------------------------------------
\section{Introduction} 

The visible Universe is made up of matter. This statement is mainly based on
observations of the cosmic diffuse $\gamma$-ray background, which
would be larger than the present limits if boundaries between
``worlds'' and ``anti-worlds'' existed \cite{cohen98}.
The observed baryon asymmetry of the universe was eventually determined
at the EWPT \cite{KRS85}. This phase transition was the
last instance when baryon asymmetry could have been generated around
$T \! \approx \! 100\!-\!200$ GeV.
Also at these temperatures any B+L asymmetry
could have been washed out. The possibility of baryogenesis at
the EWPT is particularly attractive.

Peturbation theory (PT) does not give reliable EWPT predictions for 
larger Higgs
boson masses \cite{4d_pert,3d_pert} in the standard model (SM).
 Large scale numerical simulations
both on 4d and 3d lattices were needed to analyze the nature of
the transition \cite{4d_latt,3d_latt}. They predict  \cite{3d_end,4d_end}
an end point for the first order EWPT at Higgs boson
mass 72.1$\pm$1.4 GeV \cite{4d_end}.
\footnote{Note, that the perfect agreement between the 3d and 4d results 
is a nonperturbative indication that the dimensional reduction program
is correct for hard Matsubara modes.}
The present experimental lower limit of the SM Higgs boson mass,
which is well above 100~GeV, excludes any EWPT in the SM.
In order to explain the observed baryon asymmetry, extended
versions of the SM are necessary.  According to perturbative predictions the
EWPT could be much stronger in the minimal supersymmetric extension of the
standard model (MSSM) than in the SM~\cite{mssm_pert},
in particular if the stop mass is smaller than the top mass
\cite{light_stop} and at the two-loop level.
A reduced 3d version of the MSSM has recently been studied on the
lattice \cite{mssm_3d}. 
 The results show that the EWPT can
be strong enough, i.e.\ $v/T_c \! > \! 1$, up to
$m_h \! \approx \! 105$ GeV and $m_{\tilde t} \! \approx \! 165$ GeV.
The possibility of spontaneous CP violation for a successful baryogenesis
is also addresed \cite{spont_CP}.
In this talk we review our study \cite{c_mssm} of the EWPT in the MSSM 
on 4d lattices.

\section{MSSM 4d simulation}

Except for the
U(1) sector and scalars
with small Yukawa couplings, the whole bosonic sector of the MSSM is
kept.  Fermions, owing to their heavy Matsubara modes, are
 included perturbatively in the final result.
 Our  work extends the 3d study \cite{mssm_3d} in several ways:
 \\
 a) We use 4d lattices instead of 3d. 
 Note, that due to very soft modes---close
 to the end point in the SM---much more
 CPU time is needed in 4d than in 3d.
 However, this difficulty does not appear in the MSSM because the phase
 transition is strong and the dominant correlation lengths are
 not that large in units of $T_c^{-1}$.
 Using unimproved lattice
 actions the leading corrections due
 to the finite lattice spacings are proportional to $a$ in 3d and
 only to $a^2$ in 4d. \\
 b) We  have direct
 control over zero temperature renormalization effects.
 \\
 c) We include both Higgs doublets  not only the light combination.
 According to standard baryogenesis scenarios (see e.g. \cite{mssm_gen})
 the generated baryon number is directly proportional to the change of
 $\beta$ through the bubble wall: $\Delta \beta$.
 ($\tan \beta=v_2/v_1$, where $v_{1,2}$ are the expectation
 values of the two Higgses.)
 
 The continuum lagrangian of the above theory  reads
 \begin{equation}
 {\cal L}={\cal L}_g+{\cal L}_k+{\cal L}_V+{\cal L}_{sm}+{\cal L}_Y+
 {\cal L}_w+{\cal L}_s.
 \end{equation}
 The various terms correspond to the gauge part, kinetic part for Higgses and
 third generation squarks, Higgs potential, squark mass terms , Yukawa
 couplings and quartic coupling proportional to the weak and QCD gauge
 coupling squares.
 The scalar trilinear couplings have been omitted for simplicity. It is
 straightforward to obtain
 the lattice action, for which we used the standard
 Wilson plaquette, hopping and site terms.

 The parameter space of the above Lagrangian is many-dimensional.
 The experimental values were taken for the weak, strong and Yukawa couplings,
 and $\tan \beta=6$
 was used. For the bare soft breaking masses our choice was $m_{Q,D}=250$ GeV,
 $m_U$=0 GeV. 

 Our simulation techniques are similar to those used for the SU(2)-Higgs model
 \cite{4d_latt} (overrelaxation and heatbath algorithms are used
 for each scalar and gauge field).
 For a given lattice size we fix all parameters of the Lagrangian
 except the ratio of the coefficents of the $|H_2^2|$  
and $-(\ha^\dagger \tilde{\hb}+ h.c.)$ terms in the Higgs potential.
 We tune this parameter to the transition point,
 where we determine the jump of the Higgs field, the
 shape of the bubble wall, and the change of $\beta$ through the phase
 boundary.
Next  we perform $T=0$ simulations with the same parameters and determine the masses (Higgses and W)
 and couplings (weak and strong).
 Extrapolations to infinite volumes and continuum are 
 based on simulations at various
  lattice volumes and  temporal extensions 2,3,4,5, respectively. 
  Approaching the
 continuum limit, we move on an approximate line of constant physics
 (LCP).
 Our theory is  bosonic, therefore the leading corrections
 due to finite lattice spacings are assumed  to be proportional to
 $a^2$. This lattice spacing dependence is
 assumed for physical quantities in $a\longrightarrow 0$ extrapolations.

\section{Strength of the phase transition}

We compare our simulation results with perturbation theory (PT).
We used one-loop PT without applying high
temperature expansion (HTE). A specific feature was a careful treatment
of finite renormalization effects.
This type of one-loop perturbation theory is also applied to correct
the measured data to some fixed LCP quantities, which are
defined as the averages of results at different lattice spacings, (i.e.\ our
reference point, for which the most important quantity is the lightest
Higgs mass, $m_h \approx$45 GeV).

Fig.~1. 
shows the phase diagram in the
$m_U^2$--T plane. One identifies three phases. 
The phase on the left
(large negative $m_U^2$ and small stop mass) is the
``color-breaking'' (CB) phase. The phase in the upper right part is the
``symmetric'' phase, whereas the ``Higgs'' phase can be found in the
lower right part.  The line separating the symmetric and Higgs phases
is obtained from $L_t=3$ simulations, whereas the lines between these
phases and the CB one are determined by keeping the lattice
spacing fixed while increasing and decreasing the temperature by
changing $L_t$ to 2 and 4, respectively. The shaded regions indicate
the uncertainty in the critical temperatures.  The phase transition to
the CB phase is observed to be much stronger than that
between the symmetric and Higgs phases. The qualitative features of
this picture are in complete agreement with perturbative and 3d
lattice results \cite{mssm_pert,light_stop,mssm_3d}; however, our choice of
parameters does not correspond to a two-stage symmetric-Higgs phase
transition. In the
two-stage scenario there is a phase transition from the symmetric to
the CB phase at some $T_1$ and another phase
transition occurs at $T_2<T_1$ from the CB to the Higgs phase.
It has been argued \cite{cline99} that in the early universe no
two-stage phase transition took place.

\begin{figure}[t]
\centerline{\epsfxsize=6.5cm\epsfbox{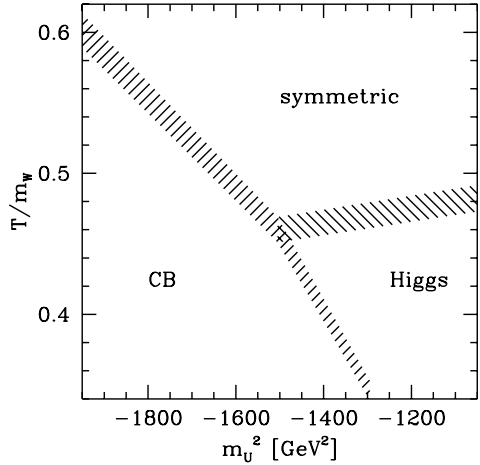}}
\caption[a]{The phase diagram of the bosonic theory obtained by lattice
simulations.}
\end{figure}

The bare squark mass parameters $m_Q^2,m_U^2,m_D^2$ receive quadratic
renormalization corrections. As it is well known, one-loop lattice
PT is not sufficient to reliably determine these corrections.
Therefore, we first determine the position of the non-perturbative CB
phase transitions in the bare quantities (e.g.\ the triple point
  or the T=0 transition for $m_U^2$ in 
  Fig.~1). 
These quantities are
  compared with the prediction of the continuum PT,
  which gives the renormalized mass parameters on the lattice.

  Fig.~2.  
shows the continuum limit extrapolation for
  the normalized jump of the order parameter ($v/T_c$: upper data)
  and the critical temperature ($T_c/m_W $: lower data). The shaded
  regions are the perturbative predictions at our reference point
  (see above) in the continuum. 
Results obtained on the lattice  and in
  PT agree reasonably within the estimated uncertainties.

\begin{figure}[t]
\centerline{\epsfxsize=6.5cm\epsfbox{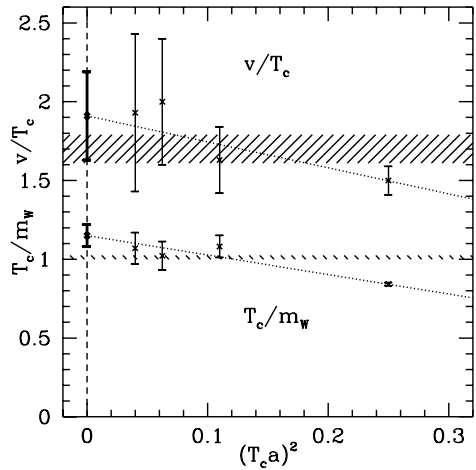}}
\caption[a]{The normalized jump and the critical temperature
in the continuum limit.}
\end{figure}

\section{Prperties of the bubble wall}

 In order to produce the observed baryon asymmetry, a strong first order
 phase transition is not enough.
  According to standard MSSM baryogenesis scenarios \cite{mssm_gen} the
  generated
  baryon asymmetry is directly proportional to the variation
  of $\beta$ through the bubble wall separating the Higgs and symmetric
  phases.
  By using elongated lattices ($2\cdot L^2 \cdot 192$), $L$=8,12,16
  at the transition point we study
  the properties of the wall. 
   In our simulation procedure
   we restrict the length of one of the Higgs fields
   to a small interval between its values in the bulk
   phases. As a consequence, the system fluctuates
   around a configuration with two bulk phases and
   two walls between them. In order to have the smallest
   possible free energy, the wall is perpendicular to the
   long direction. We eliminate the effect of the remaining
   zero mode by shifting the wall of
   each configuration to some fixed position.
  Fig.~3
  gives the bubble wall profiles for both Higgs fields.
  The measured width of the wall is
  [A+B$\cdot \log (aLT_c)]/T_c $, A=10.8$\pm$0.1 and B=2.1$\pm$0.1.
  This behavior indicates that the bubble wall is
  rough and without a pinning force of finite size its width diverges
  very slowly (logarithmically) \cite{Jasnow}.
  For the same bosonic theory the perturbative approach predicts
  $(11.2\pm1.5)/T_c$ for the width.

\begin{figure}[t]
\centerline{\epsfxsize=6.5cm\epsfbox{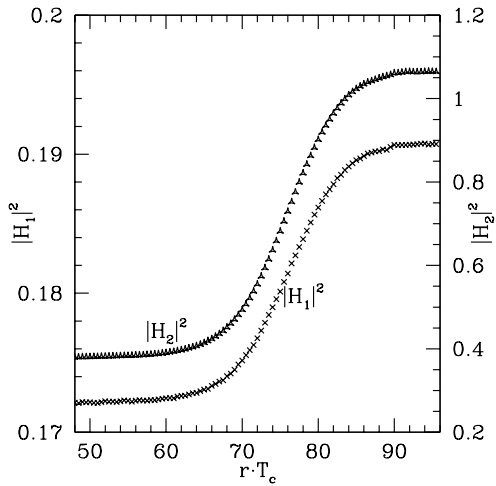}}
\caption[a]{The profile of the bubble wall for both of the Higgs fields
for the lattice $2\cdot L^2 \dot 192$.}
\end{figure}

  Transforming the data of 
  Fig.~3 
to $\vert H_2 \vert^2$ as a function of $\vert H_1
  \vert^2$,
  we obtain $\Delta \beta=0.0061 
  \pm 0.0003$.
  The perturbative prediction  is $0.0046\pm0.0010$.

\section{Conclusions}

In this talk, I have briefly discussed our work in \cite{c_mssm}. 
We presented 4d lattice results on the EWPT in the MSSM.
  Our simulations were carried out in the bosonic sector of the MSSM. We
  found quite a good agreement between lattice results and our one-loop
  perturbative predictions. We determined the phase structure of the
  MSSM and identified the three possible phases (Higgs, symmetric and
  colour broken). 
 We analyzed the bubble wall profile separating the Higgs and
  symmetric phases. The width of the wall and the change in
  $\beta$ is in fairly good agreement with perturbative predictions for
  typical bubble sizes.
  Both the strength of the phase transition and the smallness of $\Delta \beta$
  indicate that experiments allow just a small window for MSSM
  baryogenesis.
  Our results could be further checked on larger lattices, which is possible
  on a machine like PMS \cite{last}.

{\bf Acknowledgements.} This work was partially supported by
Hungarian Science Foundation Grants
OTKA-T22929-29803-M28413/FKFP-0128/1997.
The simulations were carried out on the 
46G PC-farm at E\"otv\"os University.

\vspace{1cm}

\noindent
{\small $^{\dagger}$ Presented by Z. Fodor}

\end{document}